\documentclass[superscriptaddress,nofootinbib,showpacs,preprint,preprintnumbers,amsmath,amssymb]{revtex4-1}
\usepackage{amsmath}
\usepackage{epsfig}
\usepackage{graphicx,subfigure}
\usepackage{color}
\usepackage{latexsym}
\usepackage{amssymb}
\usepackage{enumerate}

\newcommand{\gsim}{\gtrsim}
\newcommand{\lsim}{\lesssim}

\newcommand{\beq}{\begin{equation}}
\newcommand{\eeq}{\end{equation}}
\newcommand{\bea}{\begin{eqnarray}}
\newcommand{\eea}{\end{eqnarray}}
\newcommand{\barr}{\begin{array}}
\newcommand{\earr}{\end{array}}
\newcommand{\bc}{\begin{center}}
\newcommand{\ec}{\end{center}}
\newcommand{\bit}{\begin{itemize}}
\newcommand{\eit}{\end{itemize}}
\newcommand{\ben}{\begin{enumerate}}
\newcommand{\een}{\end{enumerate}}

\newcommand{\no}{\nonumber}

\newcommand{\sm}{{\rm SM}}
\newcommand{\br}{{\rm B}}

\newcommand{\pb}{{\,{\rm pb}}}
\newcommand{\fb}{{\,{\rm fb}}}

\newcommand{\gev}{{\;{\rm GeV}}}

\newcommand{\ga}{\gamma}
\newcommand{\sg}{\sigma}
\newcommand{\gm}{\gamma}
\newcommand{\Gm}{\Gamma}

\newcommand{\tb}{t_\beta}
\newcommand{\cb}{c_\beta}
\renewcommand{\sb}{s_\beta}
\newcommand{\ca}{c_\alpha}
\newcommand{\sa}{s_\alpha}

\newcommand{\cba}{c_{\beta-\alpha}}
\newcommand{\sba}{s_{\beta-\alpha}}
\newcommand{\yh}{\hat{y}}

\newcommand{\yyfu}{Y_{f_U}}
\newcommand{\yyfd}{Y_{f_D}}
\newcommand{\yfu}{y_{f_U}}
\newcommand{\yfd}{y_{f_D}}

\newcommand{\rr}{{\gamma \gamma}}
\newcommand{\ttop}{{t\bar{t}}}

\newcommand{\bb}{{b\bar{b}}}
\newcommand{\ttau}{{\tau^+ \tau^-}}

\def\fb{\,{\rm fb}}

\def\pb{\,{\rm pb}}

\def\rr{\gamma\gamma}

\def\tb{t_\beta}

\begin{document}
\baselineskip 3.5ex

\title{
Top-phobic heavy Higgs boson \\
as the 750\,GeV diphoton resonance
}

\author{Sin Kyu Kang}
\email{skkang@snut.ac.kr}
\affiliation{School of Liberal Arts, Seoul-Tech, Seoul 139-743, Korea\vspace{0.5cm}}

\author{Jeonghyeon Song}
\email{jeonghyeon.song@gmail.com}
\affiliation{School of Physics, KonKuk University, Seoul 143-701, Korea\vspace{0.5cm}}

\begin{abstract}
A hint of a new resonance at a mass of 750\,GeV has been observed in the 
diphoton channel of LHC Run 2 at $\sqrt{s}=13$ TeV. 
The signal rate is too large to interpret it as a new 
Higgs boson
in the context of weakly-coupled renormalizable models.
One way is to reduce its total decay rate,
which is possible if the CP-even heavy Higgs boson $H^0$ 
in the aligned two Higgs doublet model becomes top-phobic.
To ensure sufficient gluon fusion production,
we introduce vector-like quarks (VLQ).
The Higgs precision data as well as the exclusion limits from no excesses
in other 8 TeV LHC searches of $Z\gm$, $\bb$, $\tau^+\tau^-$, and $jj$ channels 
are simultaneously included.
In Type I, top-phobic $H^0$ cannot explain the 750\,GeV diphoton signal
and the Higgs precision data simultaneously
since the universal Yukawa couplings of the up-type and down-type 
VLQs
always make more contribution to $h^0$ than to $H^0$.
In Type II, small Yukawa coupling of the up-type VLQ but sizable Yukawa coupling
for the down-type VLQ is shown to explain the signal while satisfying other LHC exclusion limits.
\end{abstract}

\maketitle

\section{Introduction}
Very recently the ATLAS \cite{ATLAS} and CMS \cite{CMS} collaborations
have announced hints of a new diphoton resonance at a mass of 750\,GeV,
based on the 3.2\,$\mbox{fb}^{-1}$ and 2.6\,$\mbox{fb}^{-1}$ data
at $\sqrt{s}=13$\,TeV respectively.
The local significance of the diphoton excess
is 3.6$\sigma$ for the ATLAS and 2.6$\sigma$ for the CMS.
Including the look-elsewhere effect between 500\,GeV and 4\,TeV in the ATLAS data 
and between 200\,GeV and 2\,TeV in the CMS data, 
a global significance becomes less than $2.3\sigma$ for the ATLAS and $2\sigma$ 
for the CMS.
The observed excesses in the diphoton invariant mass spectrum correspond to the production cross section times branching ratio of $2.4-4.8 \fb$
according to the total decay width of the resonance particle.
Two signal rates of the ATLAS and CMS are compatible with each other.

While the excess of the data can yet be regarded as a statistical fluctuation 
which would be gone away with more data, it can be really the signal of a new particle.
In this study, we explore a possibility 
that this diphoton excess is due to a decay of a new boson.
In the literature,
there are many studies in this direction~\cite{Mambrini:2015wyu,DiChiara:2015vdm,Falkowski:2015swt,Angelescu:2015uiz,Buttazzo:2015txu,Pilaftsis:2015ycr,Franceschini:2015kwy,Ellis:2015oso,Gupta:2015zzs,Higaki:2015jag, McDermott:2015sck,Low:2015qep,Petersson:2015mkr,Dutta:2015wqh,Cao:2015pto,Kobakhidze:2015ldh,Cox:2015ckc,Martinez:2015kmn,Becirevic:2015fmu,No:2015bsn,Demidov:2015zqn,Chao:2015ttq,Fichet:2015vvy,Curtin:2015jcv,Bian:2015kjt,Chakrabortty:2015hff,Csaki:2015vek,Bai:2015nbs,Benbrik:2015fyz,Kim:2015ron,Gabrielli:2015dhk,Alves:2015jgx,Carpenter:2015ucu,Bernon:2015abk,Chao:2015nsm,Han:2015cty,Dhuria:2015ufo,Han:2015dlp,Luo:2015yio,Chang:2015sdy,Bardhan:2015hcr,Feng:2015wil,Cho:2015nxy,Barducci:2015gtd,Chakraborty:2015jvs,Han:2015qqj,Antipin:2015kgh,Wang:2015kuj,Cao:2015twy,Huang:2015evq,Heckman:2015kqk,Dhuria:2015ufo,Bi:2015uqd,Kim:2015ksf,Cline:2015msi,Bauer:2015boy,Chala:2015cev,Boucenna:2015pav,deBlas:2015hlv,Murphy:2015kag,Hernandez:2015ywg,Dey:2015bur,Huang:2015rkj,Patel:2015ulo,Chakraborty:2015gyj,Altmannshofer:2015xfo,Gu:2015lxj, Cvetic:2015vit,Allanach:2015ixl,Cheung:2015cug,Liu:2015yec,Hall:2015xds,Chang:2015bzc,Chao:2015nac,Han:2015yjk,Cai:2015hzc,Harigaya:2015ezk,Cao:2015scs,Ahmed:2015uqt,Molinaro:2015cwg,Bellazzini:2015nxw,Nakai:2015ptz,Das:2015enc,Backovic:2015fnp,Dev:2015isx,Dev:2015vjd,Bizot:2015qqo,Knapen:2015dap,Badziak:2015zez,Arun:2015ubr,Zhang:2015uuo}.
By virtue of the Landau-Yang theorem~\cite{landau-yang}, 
the new boson cannot be a spin-1 particle. 
A spin-2 particle such as a heavy graviton is unlikely 
because it has universal couplings proportional to the energy-momentum tensor,
and should have left
similar excesses in other final states like $\ell^+ \ell^-$.
Therefore we consider the best possibility that the resonance is a
spin-0 particle.

A strong candidate for the new spin-0 particle is 
a heavy neutral Higgs boson coming from physics beyond the Standard Model (SM) 
such as two Higgs doublet models (2HDM) and minimal supersymmetric SM.
However the usual heavy Higgs boson
in the context of weakly-coupled renormalizable models
has a difficulty in explaining the unexpectedly large diphoton signal rate.
For example, usual heavy Higgs bosons $H^0$ and $A^0$ in the 2HDM 
require more than three copies of vector-like quarks (VLQs)
with exotically high electric charges to explain the observed 
$\sigma\cdot \mbox{B}$~\cite{Angelescu:2015uiz}.
There are three ways to enhance the signal rate:
(\textit{i}) increasing the production cross section,
(\textit{ii}) increasing the diphoton decay rate, and/or (\textit{iii}) decreasing the total decay rate.
Methods (\textit{i}) and (\textit{ii}) 
have been extensively studied in the literature for various new physics models.
However the method (\textit{iii}) has not been much focused yet,
which is actually one competent way to obtain the relatively large diphoton signal.

We shall show that this possibility can be naturally achieved
in one of the most popular new physics models,
the aligned 2HDM.
Among five physical Higgs boson degrees of freedom 
(the light CP-even scalar $h^0$,
the heavy CP-even scalar $H^0$, the CP-odd pseudoscalar $A^0$,
and two charged Higgs bosons $H^\pm$),
we consider the case that $H^0$ is the 750\,GeV state
and $h^0$ is the observed Higgs boson.
In the alignment limit, $h^0$
has the same couplings as the SM Higgs boson.
The sum rule prohibits the couplings of
$H^0$ with $VV$ ($V=W^\pm,Z^0$).
The dominant decay channel becomes into $\ttop$,
and the diphoton branching ratio is still very small even without the $VV$ mode:
$\br(H^0 \to \rr) \sim \mathcal{O}(10^{-6})$
if the $H^0$-$t$-$\bar{t}$ coupling is SM-like.
But this $\ttop$ decay mode can be suppressed by increasing the $\tan\beta$
parameter since
 the top Yukawa coupling of $H^0$ in the aligned 2HDM
is inversely proportional to $\tan\beta$.
Then $H^0$ becomes top-phobic.
In what follows, 
the top-phobic Higgs boson is meant by $H^0$ 
with the partial decay rate $\Gamma(H^0 \to \ttop)$
below one percent of the SM value.

If there exist only SM fermions, however,
small $H$-$t$-$\bar{t}$ vertex necessarily suppresses the gluon fusion production
which occurs radiatively through the top quark loop.
New colored particles are required. 
We introduce vector-like quarks (VLQs) whose couplings to $H^0$ play a dominant role 
in the gluon fusion
production as well as the loop-induced decay 
to diphoton~\cite{Bonne:2012im}.
Since the same VLQs also contribute to the loop-induced couplings  of $h^0$ 
to $gg$ and $\rr$,
a consistency check with the Higgs precision data is 
necessary.
Moreover 
one should consider simultaneously 
that no excesses for the comparable mass scale have been observed 
in any other channels like $Z\gm$, $\bb$, $\tau^+\tau^-$, $jj$, $ZZ$, and $W^+ W^-$.
So our main question is whether the top-phobic $H^0$ 
in the aligned 2HDM can be the observed 750\,GeV state
while satisfying these LHC data. 
In addition, we also study the phenomenological characteristics of this top-phobic
$H^0$ at the 13\,TeV LHC.

The paper is organized as follows.
We describe 
the top-phobic heavy Higgs boson of the aligned 2HDM
in Sec.~\ref{sec:review}.
We introduce VLQs and demonstrate their contributions to the effective couplings of 
$H^0/h^0$ to $gg$ and $\ga\ga$.
The gain due to the \emph{top-phobic} nature of $H^0$
is to be also discussed.
In Sec.~\ref{sec:numerics}, 
we perform a parameter scan to account for the signal rate while
imposing various constraints.
The numerical results on the final allowed parameter space shall be presented.
In Sec.~\ref{sec:conclusions}, we draw our conclusions. 

\section{top-phobic heavy Higgs boson in the aligned 2HDM with VLQ}
\label{sec:review}
\subsection{Review of 2HDM and top-phobic $H^0$}
We consider a 2HDM with CP invariance
and softly broken $Z_2$ symmetry~\cite{2HDM}.
There are two Higgs doublet fields, $\Phi_1$ and $\Phi_2$,
which develop nonzero vacuum expectation values (VEVs) $v_1$ and $v_2$ respectively.
There are five physical Higgs boson degrees of freedom:
the light CP-even scalar $h^0$,
the heavy CP-even scalar $H^0$, the CP-odd pseudoscalar $A^0$,
and two charged Higgs bosons $H^\pm$.
The SM Higgs field
is a mixture of $h^0$ and $H^0$ as
\bea
H^{\rm SM} = \sba h^0 + \cba H^0,
\eea
where we take $s_x=\sin x$, $c_x = \cos x$, and $t_x = \tan x$ for simplicity of notation,
$\alpha$ is the mixing angle between $h^0$ and $H^0$,
and $\tb = v_2/v_1$.
Although the current LHC Higgs precision data can also be explained by $H^0$~\cite{Wang:2014lta,Kanemura:2014dea,Bernon:2014nxa,Coleppa:2013dya,deVisscher:2009zb,Ferreira:2014dya,Chang:2015goa,Bernon:2015wef}, the observed 125 GeV state is set to be $h^0$.
We take the alignment limit~\cite{alignment}, $\sba=1$, 
so that $h^0$ has the same couplings as the SM Higgs boson~\cite{2hdm:Higgs:fit}.


We consider the case where the 750\,GeV state is the heavy CP-even
Higgs boson $H^0$:
\bea
M_H =750\gev .
\eea
We also assume that the pseudoscalar $A^0$ and the charged Higgs $H^{\pm}$ are so heavy 
that they do not affect the neutral Higgs decays and productions.
In this study we do not consider the $H^0$-$h^0$-$h^0$ coupling,
although it can impose many interesting implications.
In the alignment limit, the $H^0$-$V$-$V$ ($V=W^\pm,Z^0$) couplings vanish.
The Yukawa couplings are different according to the $Z_2$ charges of the SM fermions,
which determine the types of 2HDM.
In Type I and Type II, the normalized Yukawa couplings by the SM values 
are
\bea
\label{eq:Yukawa:couplings:H}
\renewcommand*{\arraystretch}{1.2}
\hbox{Type I: }  && ~~\yh^H_{t} = \yh^H_{b}=\yh^H_{\tau} =-\frac{1}{\tb},  \\ 
\nonumber
\hbox{Type II: } & & ~~\yh^H_{t} =-\frac{1}{\tb}, \quad \yh^H_{b}=\yh^H_{\tau}=\tb.
\eea
In both Type I and Type II,
large $\tb$ yields the top-phobic $H^0$,
for which we require $\Gamma(H^0 \to \ttop)/\Gamma(H^0_{\rm SM} \to \ttop) \leq 1\%$.
Note that large $\tb$ leads to small $\alpha$ in the alignment limit.
In Type II, $\yh^H_b$ and $\yh^H_\tau$
are proportional to $\tb$:
too large $\tb$ 
is to be excluded by the $\tau^+\tau^-$ resonance searches 
at the 8 TeV LHC~\cite{tautau,Song:2014lua}.

\subsection{Contributions of Vector-like Quarks}
To provide sufficient gluon fusion production of the top-phobic $H^0$, 
we need new contributions to the triangular loops for the $H^0$-$g$-$g$ vertex. 
 For the purpose, we take into account extra VLQs consisting of $Q_{L/R}$, 
 $U_{L/R}$, and $D_{L/R}$.
The $SU(3)_c \times SU(2)_L \times U(1)_Y$ quantum numbers of $Q_{L/R}, U_{L/R}, D_{L/R}$ are
 $(\mathbf{3}, \mathbf{2}, -5/3)$,  $(\mathbf{3}, \mathbf{1}, -2/3)$,
 and  $ (\mathbf{3}, \mathbf{1}, -8/3)$, respectively.
The $SU(2)$ doublet $Q_{L/R}$ is presented by
\begin{eqnarray}
Q_{L/R}=\left( \begin{array}{c}
              U^{\prime} \\
              D^{\prime} \end{array} \right)_{L/R}.
\end{eqnarray}
The interactions of VLQs are described by
\begin{eqnarray}
-\mathcal{L}=&&Y_{f_L^d}\bar{Q}_L D_R H_d
+ Y_{f_R^d}\bar{Q}_{R}D_L H_d+
Y_{f_L^u} \bar{Q}_L U_R \tilde{H}_u
+ Y_{f_R^u} \bar{Q}_R U_L \tilde{H}_u \nonumber \\
&&+m_Q \bar{Q}_L Q_R-m_U\bar{U}_L U_R-m_D\bar{D}_L D_R+h.c.,
\label{VLQs}
\end{eqnarray}
where $\tilde{H}_{u}=i\tau_2H_{u}^{\ast}$.
In Type I, $H_u =H_d = \Phi_2$,
and in Type II, $H_u =\Phi_2$ and $H_d = \Phi_1$.
For simplicity, we assume
$Y_{f^u_L}=Y_{f^u_R}=\yyfu$ and $Y_{f^d_L}=Y_{f^d_R}=\yyfd$.
In addition, there exist Yukawa interactions such as 
$(\bar{q}_L U_R H_u+h.c.)$ in Type I and Type II as well as
$(\bar{Q}_L d_R \tilde{H}_d+h.c)$ in  Type II, 
where $q_L$ and $d_R$ are the SM quark doublet and down-type quark singlet respectively. 
When the two Higgs doublet fields acquire VEVs, 
those terms give rise to mixing between $d_L$ and $U^{\prime}_L$, and $d_R$ and $U_R$,
respectively. 
Flavor changing neutral currents coming from various experiments 
constrain mixing parameters and coupling constants.
It is beyond the scope of this work to study those constraints in detail, and thus we simply assume that they are small enough.

There exist upper bounds on the masses of VLQs 
from the direct searches at the Tevatron and LHC.
If the main decay mode of VLQs includes the third generation quarks
such as $Vb$ and $Vt$, 
the mass bounds for VLQs are rather strong: $M_{\rm VLQ}\gsim 400-600$ GeV~\cite{Okada}.
If VLQs mix only with lighter generations, the mass bounds become less than 400\,GeV \cite{Okada}.
In what follows, we take the lighter mass of VLQs to be around 400\,GeV,
which is possible by
assuming that the VLQs dominantly decay into light quarks.

After the two Higgs doublet fields acquire VEVs, 
the mass matrices of the VLQs are given by
\begin{eqnarray}
\label{eq:VLQ:mass:matrix}
\mathcal{M}_U=\left ( \begin{array}{cc}
            m_Q & \frac{\yyfu }{\sqrt{2}}v_u \\
           \frac{\yyfu}{\sqrt{2}} v_u& -m_U 
              \end{array} \right), \quad
\mathcal{M}_D=\left ( \begin{array}{cc}
            m_Q & \frac{\yyfd}{\sqrt{2}}v_d \\
           \frac{\yyfd}{\sqrt{2}}v_d & -m_D
              \end{array} \right),       
\label{MVLQs} 
\end{eqnarray}
where $v_u=v_d = v \sb$ in Type I while $v_u = v\sb$ and $v_d =v\cb$ in Type II.
Then the Yukawa couplings in the mass eigenstates $U_i$ and $D_i$ ($i=1,2$)
are
\bea
-\mathcal{L}
&=&
\left\{
\begin{array}{ll}
\left( \ca h + \sa H \right)
\left( \yfu \sum_i \bar{U}_i U_i + \yfd \sum_i \bar{D}_i D_i \right); &
\hbox{ in Type I} \\
\left( \ca h + \sa H \right) \yfu \sum_i \bar{U}_i U_i 
+
\left( -\sa h + \ca H \right)\yfd \sum_i \bar{D}_i D_i ; &
\hbox{ in Type II.}
\end{array}
\right.
\eea
Here $\yfu$ and $\yfd$ are the Yukawa couplings in the mass eigenstates, given by
\bea
\label{eq:yf}
y_{f_{U,D}} = \frac{1}{\sqrt{2}} Y_{f_{U,D}} s_{2 \theta_{U,D}},
\eea
where $\theta_{U,D}$ are the mixing angles of $\mathcal{M}_{U,D}$ in Eq.~(\ref{eq:VLQ:mass:matrix}).

The new VLQs contribute to Higgs decay rates to $\rr$ and $gg$  
at one loop level.
In order to incorporate the NNLO QCD and NLO EW corrections
to the $h^0/H^0$ production and decay,
we take the well-known SM Higgs boson results for the gluon fusion production
cross section and decay rates, and multiply them by 
the relative factor $c_{jj}^{h/H}$ ($j=g,\gm$):
\begin{eqnarray}
c^{h/H}_{jj} = \frac{\Gm(h/H \to jj)}{\Gm(H_{\rm SM} \to jj)},
\end{eqnarray}
where $M_{H_{\rm SM}}=125 \gev$ for $c^{h}_{jj}$
and $M_{H_{\rm SM}}=750 \gev$ for $c^{H}_{jj}$.

The decay rates of $h^0$ and $H^0$ into $\rr$ and $gg$ in the VLQ-2HDM
are
\bea
\no
\label{eq:Gam:rr}
\Gamma(h/H \rightarrow \gamma\gamma) 
&=&
\frac{\alpha^2 m^3_{h/H}}{256 \pi^3 v^2}
\left| \sum_{i=t,b,\tau} N_C Q_f^2 \yh^{h/H}_{i} A_{1/2}(x^{h/H}_i)+
c_V^\Phi A_1(x^{h/H}_W)  + N_C \mathcal{A}^{h/H}_{\rr}
 \right|^2,
\\ \label{eq:Gam:gg}
\Gamma({h/H} \to gg) &=& \frac{\alpha_s^2 m_{h/H}^3}{128 \pi^3 v^2}
\left|
\sum_{i=t,b} \yh^{h/H}_i A_{1/2}(x^\Phi_{i})
+ 
\mathcal{A}^{h/H}_{gg}
\right|,
\eea
where $c_V^h = \sba$, $c_V^H =\cba$, $x^k_i=(m_k/2 m_i)^2$
and the loop functions $A^H_{1/2}(x)$ and $A^H_{1}(x)$ are 
referred to Ref.~\cite{Djouadi:2005gi}.
$\mathcal{A}^{h,H}_{\rr,gg}$ 
are the VLQ contributions.
The contributions of the heavy charged Higgs bosons are ignored.

In Type I,
$\mathcal{A}^{h,H}_{\rr,gg}$ are 
\bea
\hbox{Type I: } \quad 
\mathcal{A}^h_{\rr,gg}
&=&
\ca\sum_{i}
 \left\{
 B^U_{\rr,gg} \yfu\frac{v}{m_{U_i}} A_{1/2}(x^h_{U_{i}})
+
 B^D_{\rr,gg} \yfd\frac{v}{m_{D_i}}A_{1/2}(x^h_{D_{i}})
 \right\},
 \\ \no
\mathcal{A}^H_{\rr,gg}
&=&
\sa\sum_i
 \left\{
 B^U_{\rr,gg} \yfu  \frac{v}{m_{U_i}} A_{1/2}(x^H_{U_{i}})
+
B^D_{\rr,gg} \yfd \frac{v}{m_{D_i}}A_{1/2}(x^H_{D_{i}})
 \right\},
\eea
where $B^{U,D}_{\rr} = Q_{f_{U,D}}^2$ and $B^{U,D}_{gg} = 1$.
Note that the VLQ contributions to $h^0$ and $H^0$ are the same
except for $\ca$ and $\sa$ factors.
Since $\alpha \ll  1$ for the top-phobic $H^0$ in the alignment limit,
VLQ contribution to $h^0$ is much larger than that to $H^0$.
Type I cannot explain both Higgs precision data and 750\,GeV diphoton excess, simultaneously.

In Type II, the VLQ contributions to the amplitudes are
\bea
\label{eq:A:H:rr}
 \hbox{Type II: } \quad 
 \mathcal{A}^h_{\rr,gg}
&=&
\sum_{i=1,2}
 \left\{
 \ca B^U_{\rr,gg} \yfu\frac{v}{m_{U_i}} A_{1/2}(x^h_{U_{i}})
-\sa
 B^D_{\rr,gg} \yfd\frac{v}{m_{D_i}}A_{1/2}(x^h_{D_{i}})
 \right\},
 \\ \no
\mathcal{A}^H_{\rr,gg}
&=&
\sum_{i=1,2}
 \left\{
\sa B^U_{\rr,gg} \yfu\frac{v}{m_{U_i}} A_{1/2}(x^H_{U_{i}})
+
\ca B^D_{\rr,gg} \yfd\frac{v}{m_{D_i}}A_{1/2}(x^H_{D_{i}})
 \right\}.
\eea
%
It is clearly seen that since $\alpha \ll 1$,
$\yfu$ contribution is dominant to the Higgs precision data
while $\yfd$ contribution is dominant to the 750\,GeV diphoton data. 
Large signal rate of $gg \to H \to \rr$ requires sizable $\yfd$
but  the SM-like Higgs data
do small $\yfu$. 
This feature combined with large $\tb$
has an important implication
on the VLQ mass matrices in Eq.~(\ref{eq:VLQ:mass:matrix}).
In Type II, the off-diagonal elements of $\mathcal{M}_U$ and $\mathcal{M}_D$
become smaller than the diagonal elements if we assume $m_{Q,U,D} \gsim 400\gev$.
The mixing is subdominant.

\subsection{Enhanced diphoton rate of the top-phobic $H^0$}

\begin{figure}[t]
\begin{center} 
\includegraphics[width=0.6\textwidth]{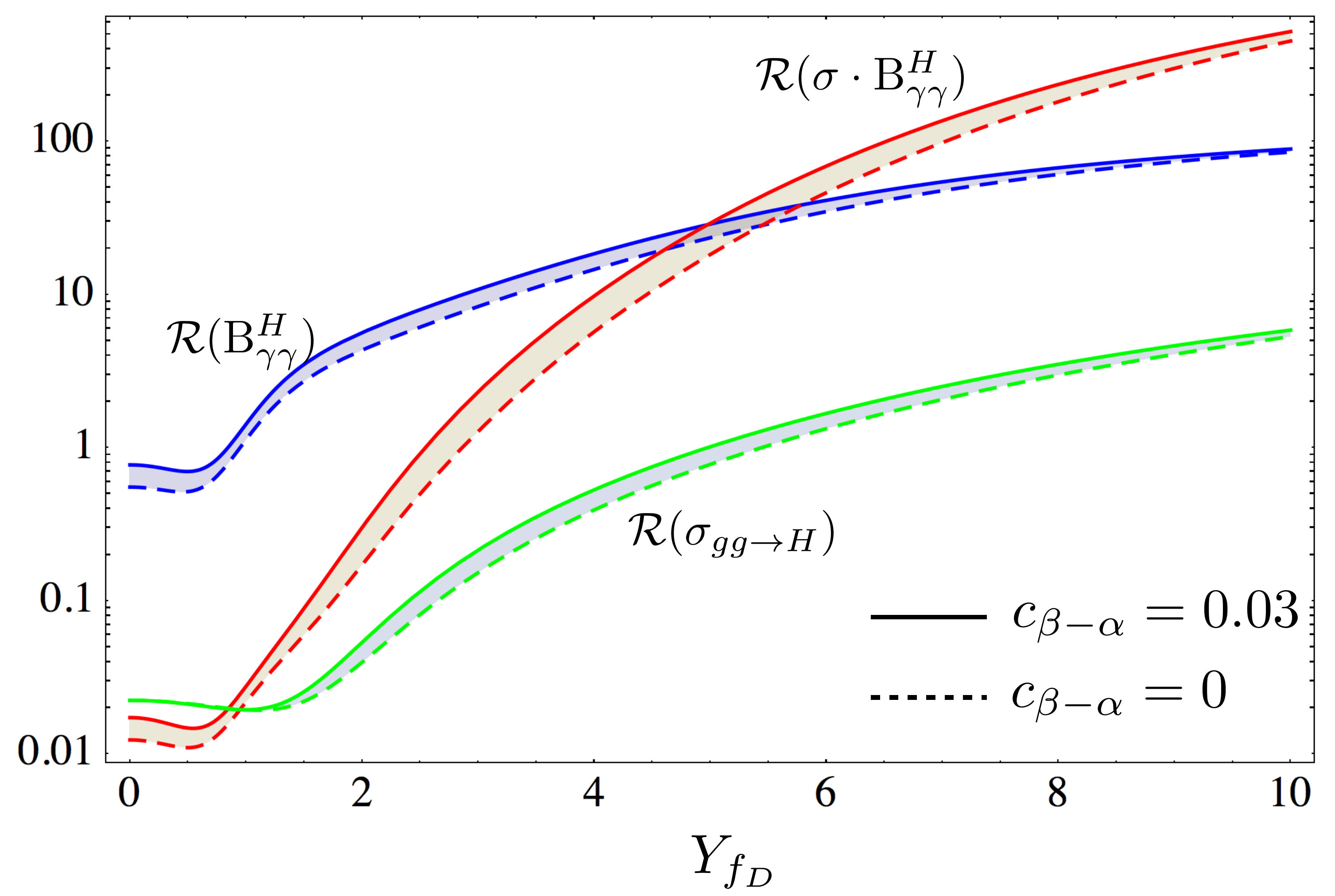}
  \caption{\baselineskip 3.5ex \label{fig:ratio}
  The diphoton branching ratio, the $gg\to H \to \rr$ signal rate,
  and the gluon fusion production cross section 
  of the top-phobic $H^0$ normalized by the SM value.
  For the $H^0$ in the VLQ-2HDM,
  we set  $m_Q = 500\gev$, $m_U = 800 \gev$, $m_D = 380\gev$, $\yyfu=0.5$,
 $\mathcal{R}(\Gamma_{H \to \ttop})=1\%$,
 and $\cba=0.03\hbox{ (solid line)}, 0\hbox{ (dashed line)}$.
   }
\end{center}
\end{figure}

One of the main reasons why the possible 750\,GeV state 
with $\sigma\cdot\br\approx 2.4-4.8\fb$
cannot be the SM-like Higgs boson
is the extremely small diphoton branching ratio of $H_{750}^{\rm SM}$~\cite{Song:2014lua}.
In order to dramatically enhance it,
we take the alignment limit and the top-phobic $H^0$,
which prohibits the decays into $VV$ and 
suppresses the decay into $\ttop$ respectively.
In Fig.~\ref{fig:ratio},
we present the gain of the top-phobic $H^0$ relative to the SM Higgs boson 
at a mass of 750\,GeV through their ratio of the diphoton branching ratio,
the signal rate, and the gluon fusion production cross section
as a function of $\yyfd$.
Here $\mathcal{R}(O) \equiv O/O_\sm$. 
We set $m_Q = 500\gev$, $m_U = 800 \gev$, $m_D = 380\gev$, $\yyfu=0.5$,
and $\mathcal{R}(\Gamma_{H \to \ttop})=1\%$.
Two values of $\cba$ are considered, $\cba=0.03$ (solid line) and $\cba=0$ (dashed line).
All of three $\mathcal{R}(\br^H_{\rr})$,
$\mathcal{R}(\sigma_{gg\to H})$, and $\mathcal{R}(\sg\cdot\br)$
have quite similar results for two $\cba$'s.
Small deviation from the alignment does not yield dramatic changes 
in the results.

It is clear to see that
$\br (H \to \rr)$
can be significantly enhanced if $\yyfd \gsim 2$.
This is mainly attributed to the suppressed decay rate into $\ttop$.
On the while, the gluon fusion production cross section 
is smaller than the SM value if $\yyfd \lsim 5$, 
since the small top Yukawa coupling of $H^0$
suppresses the $g$-$g$-$H^0$ vertex at one loop level.
When $\yyfd \gsim 5$,
the VLQ contribution becomes compatible 
with the top quark contribution for the SM Higgs boson. 
In combination, $\sigma\cdot\br$ can increase by an order of magnitude 
if $\yyfd \gsim 4$.


\section{Numerical Results}
\label{sec:numerics}

The main question of this study is whether the top-phobic $H^0$
in the aligned VLQ-2HDM
can explain the possible 750\,GeV state while satisfying the other LHC constraints.
We consider the following three classes of observations:
\begin{enumerate}
\item \textit{The diphoton resonance at 750\,GeV}:
Based on the ATLAS and CMS combined results of 8 TeV
and 13 TeV, we accept the best-fit results
by varying the total width $\Gamma$ and $\sigma\cdot \br$ through a Poissonian likelihood analysis~\cite{Falkowski:2015swt}.
We fix the new particle mass at $750\gev$.
The allowed value of $\sigma\cdot \br$
is different according to $\Gamma$.
Since the top-phobic $H^0$ in the alignment limit
has small total decay width like
$\Gamma \sim 5\gev$ we have
$\sigma\cdot \br=
2.4^{+ 1.35}_{-1.30}\fb$.
Note that if the width is large like $\Gm=30\gev$,
the best-fit value increases to be
$\sigma\cdot \br=4.8^{+2.1}_{-2.3}\fb$.

\item \textit{Higgs precision data}:
We impose the constraints from the Higgs precision data, particularly the ATLAS~\cite{Aad:2014eha} and CMS~\cite{Khachatryan:2014ira,Khachatryan:2014jba} measurements of the signal strength of $\mu^{\rm ggF}_{\rr}$: 
\bea
\mu^{\rm ggF}_{\rr} =\left\{\begin{array}{lc}
                            1.32 \pm 0.38 & (\mbox{ATLAS});  \\
                            0.85^{+0.19}_{-0.16} & (\mbox{CMS}).
                            \end{array} \right.
\eea
\item \textit{Exclusion from no observation of new resonance searches at the 8 TeV LHC}:
We consider the following upper bounds 
on the signal rate of the 750\,GeV $H^0$:
	\begin{enumerate}
	\item $\sigma(pp \to H \to Z \gamma)
	\leq 4\fb$~\cite{Zr};
	\item $\sigma(pp \to H \to \ttau)
	\leq 12\fb$ \cite{tautau}.
	\item $\sigma(pp \to H \to b\bar{b})
	\leq 1 \pb$ \cite{bb};
	\item $\sigma(pp \to H \to jj)
	\leq 12 \pb$~\cite{jj};
	\item $\sigma(pp \to H \to Z Z)
	\leq 20\fb$~\cite{ZZ}.
	\end{enumerate}
\end{enumerate}
When computing the signal rates,
we use the SM results of the gluon fusion production cross section
at $\sqrt{s}=13$ TeV of
$\sigma(H_{\rm SM})=0.85 \pb$~\cite{sigma13TeV}.
The diphoton signal rate is $c_{gg}^H \times\sigma(H_{\rm SM})\times \br(H \to \rr)$.
The NNLO QCD and NLO EW corrections are naturally included.
We do not consider the interference with the continuum background~\cite{intf}.

\begin{figure}[t]
\centering
\includegraphics[width=0.5\textwidth]{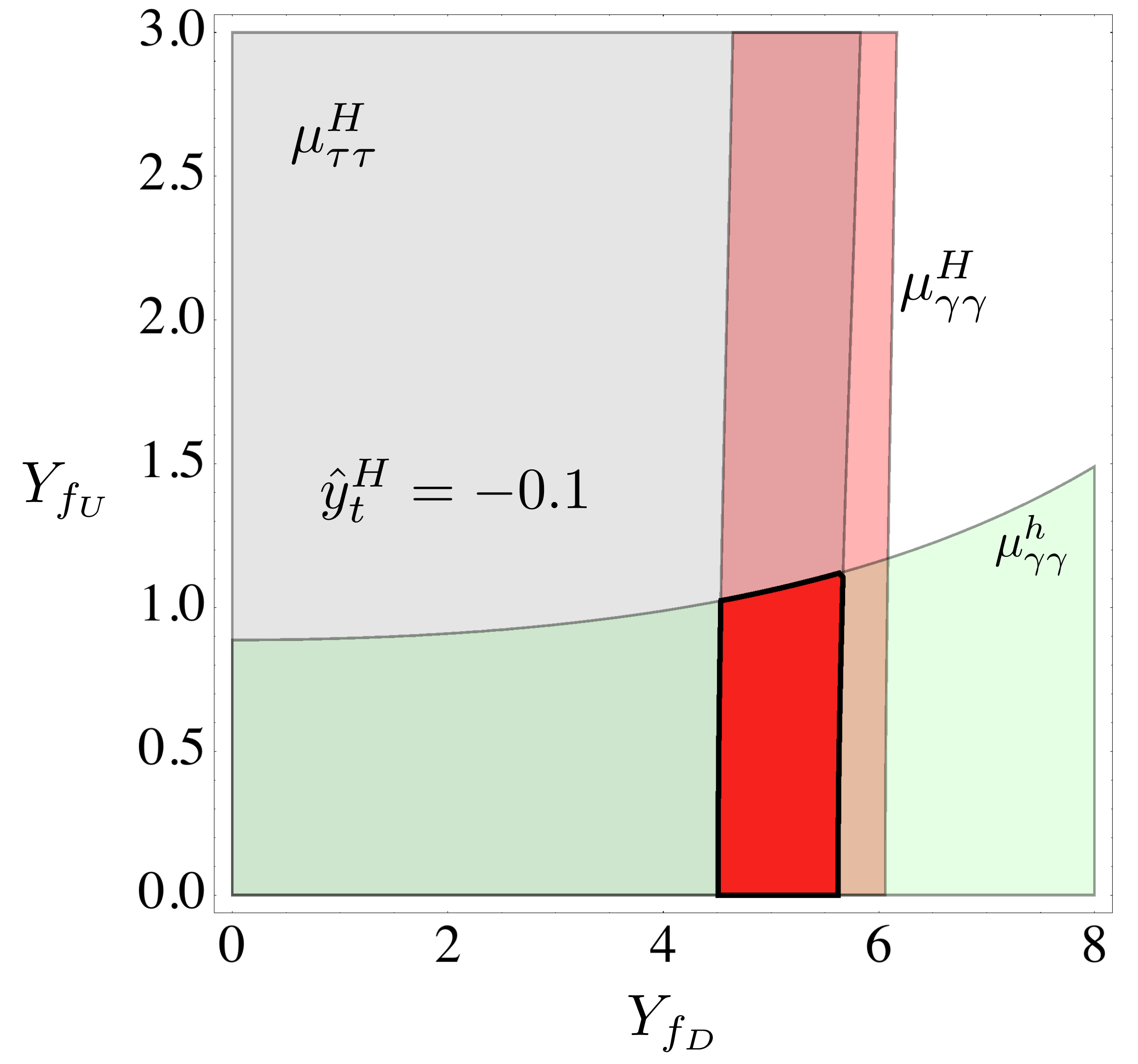}
\caption{\label{fig:allowed}
Allowed region in the parameter space $(\yyfd,\yyfu)$
in Type II VLQ-2HDM.
We have set 
$m_Q = 500\gev$, $m_U = 800 \gev$, $m_D = 380\gev$, $\cba=0$,
and $\mathcal{R}(\Gamma_{H \to \ttop})=1\%$.
The pale red region explains the 750\,GeV diphoton signal,
the green one  by the Higgs precision data,
and the grey one by 
$gg\to H \to \ttau$ at the 8 TeV LHC.
 }
\end{figure}

Figure \ref{fig:allowed} shows the 95\% C.L. 
allowed parameter space $(\yyfd,\yyfu)$
in Type II by the LHC diphoton excess at 750\,GeV (pale red region),
the Higgs precision data (green region),
and the $\ttau$ data from the 8\,TeV LHC (grey region).
We have set 
$m_Q = 500\gev$, $m_U = 800 \gev$, $m_D = 380\gev$, $\cba=0$,
and $\mathcal{R}(\Gamma_{H \to \ttop})=1\%$.
One copy of $Q_{L/R}$, $U_{L/R}$ and $D_{L/R}$ is included.
The large diphoton signal rate of the 750\,GeV state
requires sizable $\yyfd$ around 5, but remains almost insensitive to $\yyfu$.
This is attributed to small $\alpha$ in Eq.~(\ref{eq:A:H:rr}).
 
The allowed region by the Higgs precision data is denoted by the
green region. As discussed before,
the Higgs precision data is sensitive to $\yyfu$ but not to $\yyfd$.
The diphoton signals of the 750\,GeV and 125 GeV states
play complementary roles in determining $\yyfu$ and $\yyfd$:
the former fixes  $\yyfd \sim 5$ and the latter
$\yyfu\lsim 1$.
We also present the most sensitive exclusion limit from the LHC8 data,  
$gg\to H \to \ttau$, by the grey region.
Large $\yyfd$ 
enhances the gluon fusion production cross section
and large $\tb$ increases the branching ratio of $H \to \ttau$.
$\yyfd \approx 6$ is excluded.
More data in the $\ttau$ channel at the LHC13 will play a crucial role
in probing the model.
The other exclusion limits 
in the channels of $Z\gm$, $\bb$, $jj$ and $ZZ$
are all satisfied in the presented parameter space.
The final combined allowed region is the red region bounded by solid lines.

Brief comments on the perturbativity of the Yukawa couplings
are in order here.
Rather large value of $\yyfd \sim 5$ 
may cause worry about 
dangerously large contribution of the next order loop.
Note that $\yyfd$ is the Yukawa coupling in the weak basis.
What matters in the perturbation calculation
is the Yukawa couplings in the mass basis, $\yfd$.
The maximum value of $\yfd$ in the combined allowed region (bounded red region)
is much small like $\sim 0.85$.
The reduction is because of the small mixing angle $\theta_{D}$ in Eq.~(\ref{eq:yf}).
One loop level calculation for $H$-$\gm$-$\gm$ and $H$-$g$-$g$ vertices
is sufficient.

\begin{figure}[t]
\begin{center}
\includegraphics[width=0.6\textwidth]{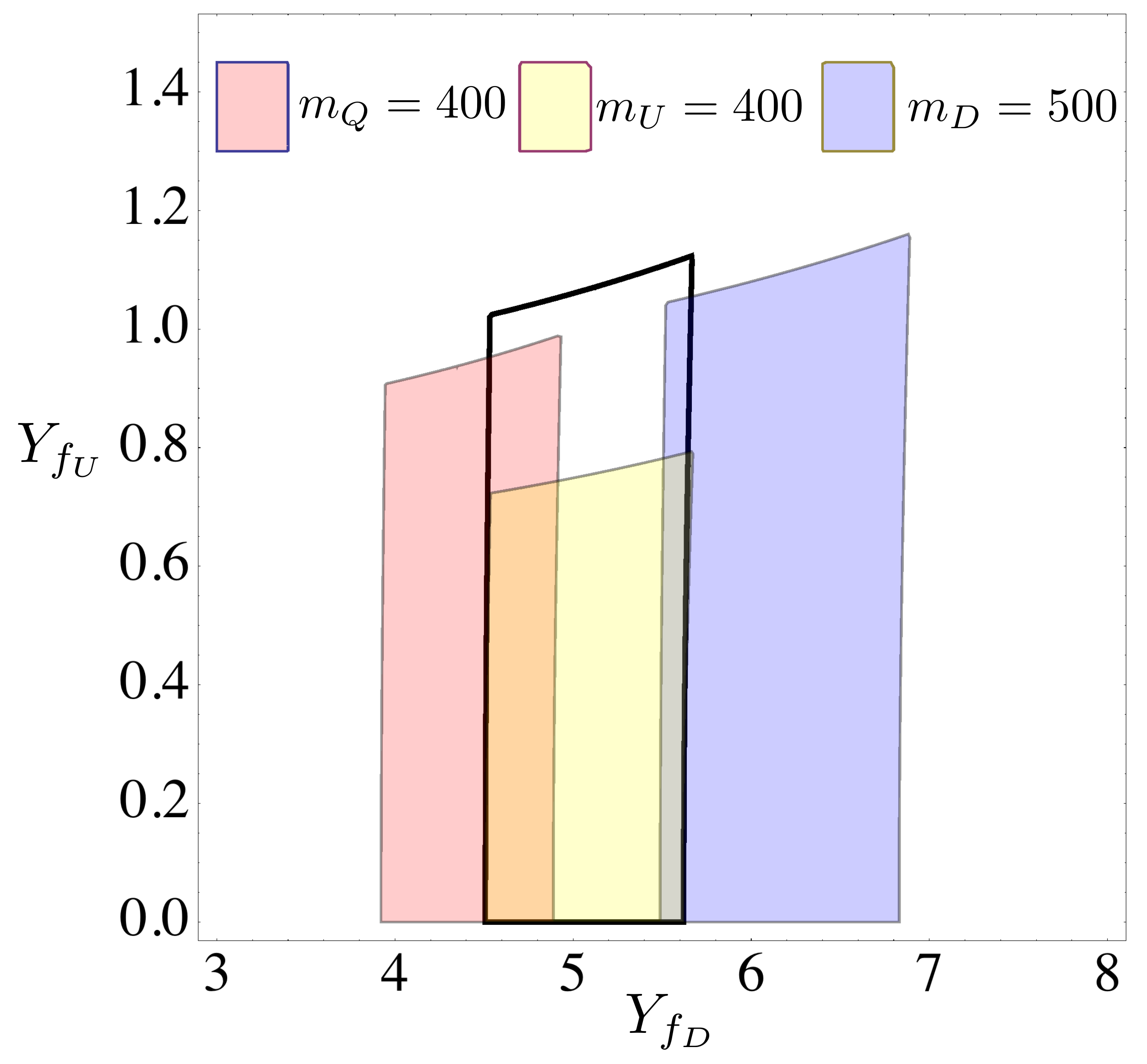}
\end{center}
\caption{\label{fig:allowed:Mass}
The final allowed region in the parameter space $(\yyfd,\yyfu)$.
The benchmark point (bounded by solid line) is for
$m_Q = 500\gev$, $m_U = 800 \gev$, $m_D = 380\gev$,
$\cba=0$,
and $\mathcal{R}(\Gamma_{H \to \ttop})=1\%$.
The red region is  for $m_Q=400\gev$
while the other parameters are the same as the benchmark point,
the yellow region for $m_U =400\gev$,
and the blue region for $m_D =500\gev$.
 }
\end{figure}

Other important model parameters are the masses of VLQs.
The benchmark point in Fig.~\ref{fig:allowed} is for
$m_Q = 500\gev$, $m_U = 800 \gev$, and $m_D = 380\gev$,
which enhances the $H^0$-$\gm$-$\gm$ vertex by setting $m_{D_{1,2}}$
near $m_H/2$ while reduces the $h^0$-$\gm$-$\gm$ vertex by rather large $m_U$.
In Fig.~\ref{fig:allowed:Mass},
we show the dependence of the VLQ masses
on the final allowed region.
The bounded region by solid lines is the 
allowed region for the benchmark point.
The red region is for smaller $m_Q=400\gev$ but the other parameters are 
the same as the benchmark point.
In this case,
both $m_{D_1}$ and $m_{D_2}$
are near the threshold $m_H/2$,
enhancing the amplitude 
relative to the case significantly below or above the threshold.
Smaller $\yyfd$ can explain the 750\,GeV diphoton signal.
The yellow region is for smaller $m_U =400\gev$ but the same $m_{Q,D}$.
The up-type VLQs do not affect the 750\,GeV diphoton signal, 
which has almost the same allowed value of $\yyfd$ for lighter $m_U$.
On the contrary, the Higgs precision data get 
larger contribution and thus smaller $\yyfu (\lsim 0.7)$ is required.
The blue region is for larger $m_D =500\gev$:
$\yyfd \approx 6$ is needed.
In summary, the VLQ mass dependence is not strong unless the VLQ masses
are much heavier than 500\,GeV.

\begin{figure}[t]
\begin{center}
\includegraphics[width=0.6\textwidth]{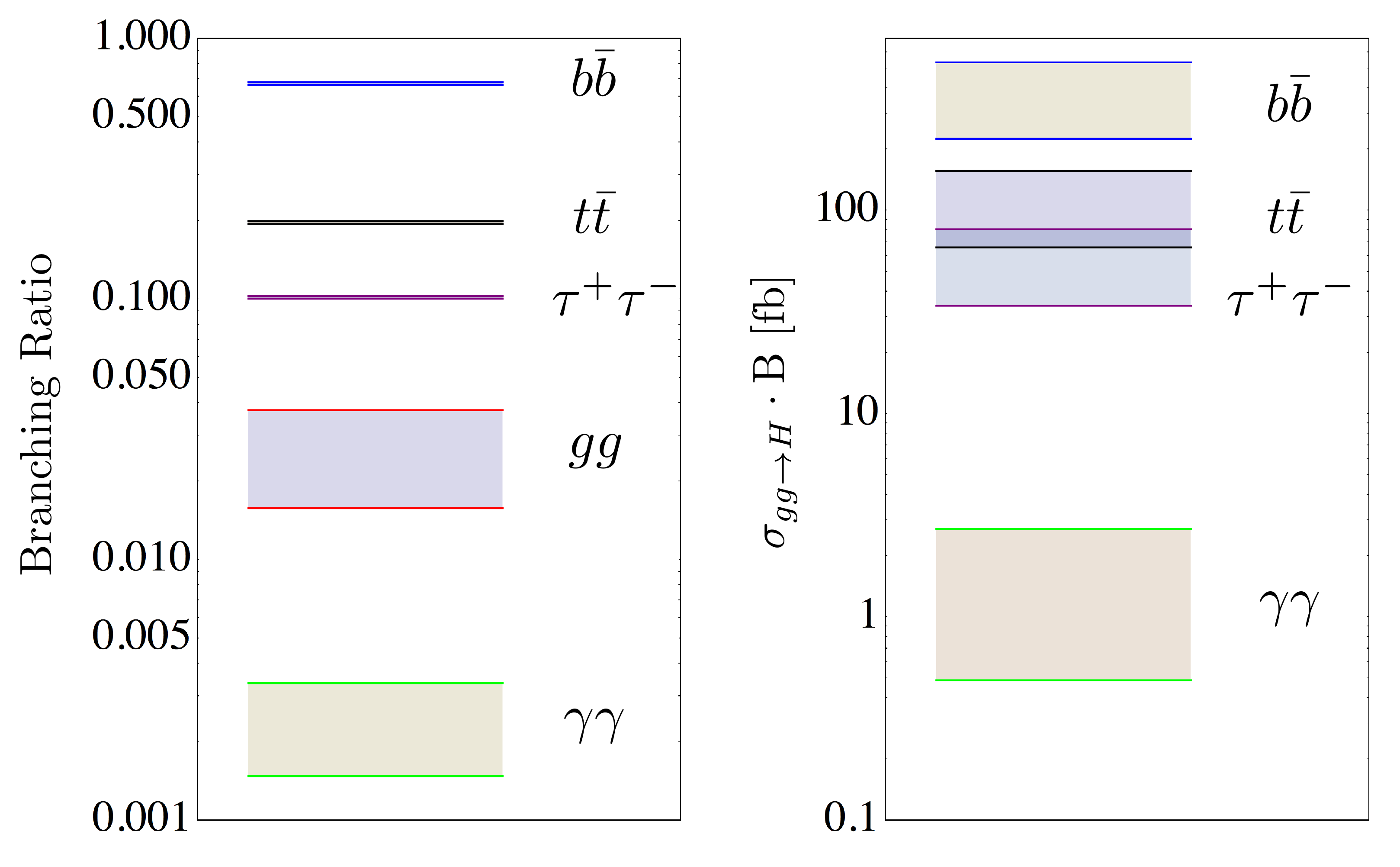}
\end{center}
\caption{\label{fig:BRs}
The branching ratios of $H^0$ (left panel) and the LHC 13 prospect
for $\sigma\cdot\br$ (right panel). 
We take $m_Q = 500\gev$, $m_U = 800 \gev$, $m_D = 380\gev$, $\cba=0$
and
$\mathcal{R}(\Gamma_{H \to \ttop})=1\%$.
For $\yyfu$ and $\yyfd$, we take the values in the final allowed region of
Fig.~\ref{fig:allowed}.
 }
\end{figure}

Finally we present the branching ratios of the top-phobic $H^0$ (left panel)
as well as the LHC 13 prospect of various signal rates (right panel)
in Fig.~\ref{fig:BRs}.
The parameter setting is the same as in Fig.~\ref{fig:allowed}.
For $\yyfu$ and $\yyfd$,
we take the values of the final allowed regions
and show the maximum and minimum values for each observable.
The diphoton branching ratio is highly enhanced,
of the order of $10^{-3}$.
The dominant decay mode is into $\bb$
since the $b$ quark Yukawa coupling is enhanced by $\tb$.
The next dominant mode is $\ttop$:
the SM top Yukawa coupling itself is large.
The third important mode is the $\ttau$ channel.

The LHC 13 prospects on $\sigma (gg \to H) \cdot \br$ in the $\bb$, $\ttop$, $\ttau$
and $\rr$ channels for the top-phobic $H^0$ are presented in the right panel.
As can be seen from the branching ratio, 
the $\bb$ has the largest signal rate of about 400\,fb
and the $\ttop$ has the second largest rate $\sim 110\fb$.
However huge QCD backgrounds shall make it difficult to measure the signal in these
hadronic channels.
The $\ttau$ signal rate about 60\,fb 
is very promising at the 13\,TeV LHC.

\section{Conclusion}
\label{sec:conclusions}
A hint of a new resonance at a mass of 750\,GeV 
has been observed in the diphoton channel of LHC Run 2 at $\sqrt{s}=13$ TeV. 
We have investigated if a top-phobic heavy neutral Higgs boson 
in the aligned two Higgs doublet model 
can be responsible for the diphoton excess. 
The relatively large signal rate observed at 13 TeV 
is efficiently accounted for 
by reducing the total decay width
and thus increasing the diphoton branching ratio. 
One good example is the top-phobic $H^0$ in the aligned 2HDM.
We also introduced vector-like quarks so that their couplings to the top-phobic Higgs 
guarantee sufficient gluon fusion production.
We have showed that in Type I the top-phobic $H^0$ cannot 
explain the 750\,GeV diphoton signal rate
since the universal Yukawa couplings of up-type ($\yyfu$) and down-type ($\yyfd$)
VLQs
always yield more contribution to $h^0$ than to $H^0$.
In Type II, we found that $\yyfu$ mainly contributes to $h^0$
while $\yyfd$ to $H^0$.
There exists the allowed parameter region of $\yyfu \lsim 1$
and $\yyfd \sim 5$
which explains 
the 750\,GeV diphoton excess 
as well as the Higgs precision data
and the exclusion limits from the 8 TeV LHC searches 
in the $Z\gm$, $\bb$, $\tau^+\tau^-$, $jj$, $W^+ W^-$, and $ZZ$ channels.
The $\tau^+\tau^-$ resonance searches at the 8 TeV LHC 
begin to constrain the model.
The dependence of the VLQ masses was shown to be moderate.

\begin{acknowledgements}
The work of SKK is supported by NRF-2014R1A1A2057665, and
the work of JS is supported by NRF-2013R1A1A2061331.
\end{acknowledgements}


\end{document}